\documentclass[twocolumn,showpacs,aps,floatfix]{revtex4}
\usepackage{amsmath}
\usepackage{graphicx}
\usepackage{dcolumn}
\usepackage{bm}

\begin{document}

\title{NMR detection with an atomic magnetometer}
\author{I. M. Savukov and M. V. Romalis}

\affiliation{Department of Physics, Princeton University,
Princeton, New Jersey 08544}

\date{\today}

\begin{abstract}
We demonstrate detection of NMR signals using a non-cryogenic
atomic magnetometer and describe several novel applications of
this technique. A water free induction decay (FID) signal in a 0.5
$\mu$T field is detected using a spin-exchange-relaxation-free K
magnetometer and the possibility of using a multi-channel
magnetometer for 3-D MRI requiring only a single FID signal is
described. We also demonstrate detection of less than $10^{13}$
$^{129}$Xe atoms whose NMR signal is enhanced by a factor of 540
due to Fermi-contact interaction with K atoms. This technique
allows detection of less than $10^{9}$ $^{129}$Xe spins in a
flowing system suitable for remote NMR applications.
\end{abstract}
 \maketitle

Nuclear magnetic resonance is a powerful technique widely used in
both basic research and medical applications. Traditionally, NMR
signals from thermal nuclear polarization are detected with an RF
pick-up coil. A high magnetic field, usually produced by a
superconducting magnet, increases the strength of the signal
approximately as $B^2$ and improves the ability to resolve NMR
chemical shifts.  In applications that do not require chemical
shift information it is possible to avoid using a large magnetic
field by utilizing a magnetometer instead of an inductive pick-up
coil to detect the NMR signal, making the signal strength
proportional only to the first power of $B$. Detection of NMR
signals in fields as low as a few $\mu$T has been demonstrated
using SQUID magnetometers \cite{Pines,Greenberg}. It is even
possible to completely eliminate the dependence of the signal
strength on the magnetic field by utilizing hyperpolarized nuclei,
such as $^{129}$Xe polarized by spin-exchange optical pumping
\cite{PinesXe} or protons polarized by Spin Polarization Induced
Nuclear Overhauser Effect \cite{Heckman}. However, NMR detection
with SQUID magnetometers still requires a cryogenic system and
precludes many applications of NMR and MRI requiring portable,
maintenance-free systems.

In this Letter we demonstrate NMR detection using an atomic
magnetometer and describe several novel applications based on
unique properties of atomic magnetometers. Recent advances in
atomic magnetometry \cite{Budker}, in particular demonstration of
a spin-exchange-relaxation-free (SERF) magnetometer \cite{Allred}
have allowed alkali-metal magnetometers to exceed the sensitivity
of low-temperature SQUID detectors \cite{Kominis}. In addition to
high sensitivity, atomic magnetometers also allow low-cost
multi-channel measurements with a high spatial resolution and do
not require cryogenic cooling. Here we demonstrate first detection
of NMR free induction decay (FID) signals from a thermally
polarized water sample with an atomic magnetometer. Previous
measurements using atomic magnetometers have only detected DC
magnetization of hyperpolarized gases
\cite{Cohen,Cates,BudkerNMR}. We also describe a novel MRI method
that allows reconstruction of a 3-D image from a single FID signal
in the presence of a constant magnetic field gradient by relying
on multi-channel magnetic field measurements. In a separate
experiment, we investigate a unique method for enhancing NMR
sensitivity by allowing the nuclei to occupy the same volume as
the active atoms of the magnetometer. In addition to reducing the
distance between the nuclear spins and the atoms measuring the
magnetic field, this method can also enhance the NMR signal due to
Fermi-contact interaction between the alkali-metal valence
electron and the nuclei spins \cite{Grover,Schaefer}.  With this
technique we detect a signal from $2\times 10^{13}$ $^{129}$Xe
atoms with a signal-to-noise ratio of $ 10$ in a single shot with
a bandwidth of 10 Hz. Straight-forward optimization of this
technique  can achieve sensitivity of about $10^9$ $^{129}$Xe
spins without averaging. For comparison, detection of NMR using
traditional pick-up coils is limited at a level $3\times 10^{12}$
spins even with substantial averaging \cite{Minard,Pennington}.
Detection of optically pumped Ga spins using magnetic resonance
force microscopy has achieved sensitivity of $7\times 10^{8}$
spins \cite{Garner}. Hyperpolarized $^{129}$Xe is widely used for
MRI \cite{Albert}, as a biosensor \cite{PinesBio, Bifone}, and for
remote NMR detection \cite{Remote}. However, measurements with
human subjects are restricted due to anesthetic properties of
xenon in high concentrations.  Combining the techniques of remote
NMR encoding with a flow-through spin-detection system using an
atomic magnetometer would allow one to use much smaller $^{129}$Xe
concentrations.

Atomic magnetometers operate by measuring the precession of
electron spins in a magnetic field, usually using an alkali metal
vapor. The sensitivity of the magnetometer is determined by the
number of atoms in the active volume and their transverse spin
relaxation time. Atomic collisions usually limit the transverse
spin relaxation time, particularly at high alkali-metal density.
As was first shown in \cite{Happer}, the dominant relaxation
mechanism due to spin-exchange collisions can be eliminated by
operating in a very low magnetic field with a high alkali-metal
density. Such spin-exchange relaxation-free (SERF) magnetometer
has achieved magnetic field sensitivity of 0.5 fT/Hz$^{1/2}$ using
an active volume of only 0.3 cm$^3$ \cite{Kominis}. The small
active volume is important for obtaining a short effective
distance between the magnetometer and the NMR sample. The
magnetometer consists of a glass cell containing K vapor and a
high pressure buffer gas to slow the diffusion of atoms across the
cell. An optical pumping laser spin-polarizes the atoms while an
orthogonal probe laser detects their precession in the magnetic
field. Because of slow K diffusion, a single probe laser expanded
to fill the whole cell can be used to simultaneously measure the
magnetic field in multiple points by imaging it on a multi-channel
photo-detector. In this arrangement most elements of the
magnetometer are common, allowing one to construct an inexpensive
system with hundreds or even thousands of channels. One challenge
for using an atomic magnetometer for NMR detection is the need to
match the resonance frequencies of the electron and nuclear spins
whose gyromagnetic ratios are different by a factor of 100-1000.
One possibility is to use a set of coils to create different
magnetic fields in the NMR sample and the magnetometer cell as was
recently demonstrated in ~\cite{BudkerNMR}. If the nuclear spins
are directly interacting with the atomic magnetometer, one can use
these interactions to match the two resonance frequencies
\cite{Kornack}.

\begin{figure}
\centerline{\includegraphics*[scale=0.45]{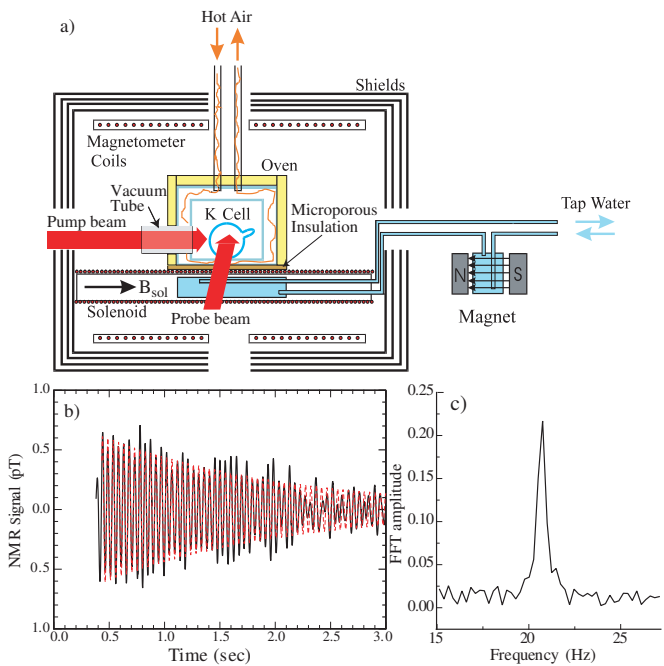}}
\caption{Water NMR Detection. a) Experimental Setup: Tap water is
thermally polarized by passing through a 2~kG permanent magnet
before flowing into a 1.25" diameter cylinder located near the
magnetometer inside magnetic shields. Pump and probe laser beams
pass through evacuated glass tubes to avoid air turbulence. The K
cell is a 1" glass sphere containing  2.5 atm of He gas and 60
torr of N$_2$ gas and a small droplet of K metal. b) Single FID
decay following a $\pi/2$ pulse. The signal is filtered with a
bandwidth of 20 Hz. From the fit (dashed line) we determine
$T_2=1.7$ sec. c) FFT of a single FID signal. The magnetic noise
has a flat spectrum with a noise floor of $2\times10^{-14}$
T/Hz$^{1/2}$. } \label{waternmr}
\end{figure}

The  experimental arrangement for detection of water NMR is shown
in Fig.~\ref{waternmr}a). To obtain independent control of the
magnetic fields experienced by the protons and the K magnetometer,
the water sample is contained in a solenoid.  The magnetic flux
produced by the solenoid is returned through the magnetic shields,
so it's external field is a factor of 1000 smaller than the
internal field. In previous experiments designed to detect NMR
with SQUIDs in a very low magnetic field the thermal polarization
was increased by briefly sending a large current through a
solenoid \cite{Pines}. In our system the inner-most magnetic
shield is made from a soft METGLAS material which easily
magnetizes and creates a large field drift. To avoid this problem
we used a flow system where the water was polarized by a permanent
magnet outside of the shields. The K cell is heated to
180$^{\circ}$C in a double-wall oven made from thin G7 sheets. Hot
air flows between the two walls of the oven but does not cross the
path of the laser beams to avoid optical noise. Microporous
thermal insulation is used to insulate the oven, keeping  the
total distance between the K cell and the room-temperature surface
to about 1 cm. Magnetometer coils inside the shields are used to
set to zero the magnetic field at the K cell in order to achieve
the maximum sensitivity of the SERF magnetometer. One coil is also
used to generate a $\pi/2$ pulse to tip the proton spins. The
transverse relaxation time of K spins is much faster than those of
protons in water, so the transient signal of the magnetometer
decays quickly relative to water FID. A single-shot water NMR
signal is shown in Fig.~1b) and its FID is shown in Fig.~1c) The
Fourier transform  reveals a single peak at a frequency of 20 Hz
 with a S/N ratio of greater than 10. The S/N ratio is
comparable or better than those obtained with SQUID magnetometers
\cite{Pines}.

\begin{figure}
\centerline{\includegraphics*[scale=0.3]{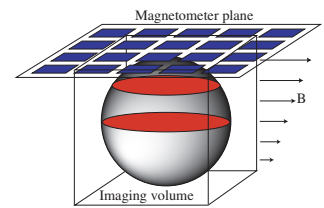}}
\caption{Schematic of an MRI technique using a planar array of
magnetometers. In a uniform magnetic field the size of a spherical
magnetization distribution can not be determined because it
produces a purely dipolar field. Applying a linear gradient
separates the object into slices in the frequency domain and the
image of each slice can be uniquely obtained from sensitive
measurements of the magnetic field outside. } \label{MRI}
\end{figure}

In addition to avoiding the need for cryogenics, NMR detection
using atomic magnetometers also allows simple construction of
multi-channel systems. The  electronics needed for each channel is
much simpler than for an RF pick-up coil or a SQUID detector and
there is no inductive coupling between different channels. This
opens the possibility of using spatial information obtained from a
large number of channels to implement more efficient MRI
techniques. Parallel MRI techniques  using phased RF arrays have
been used to reduce imaging time by omitting some phase-encoding
steps in a traditional MRI sequence \cite{Heidemann}. It is well
known that even a complete knowledge of the magnetic field outside
of a closed volume is not sufficient to reconstruct the
distribution of an arbitrary static current or magnetization
inside the volume. The situation is different in NMR, where the
magnetization starts out parallel to the magnetic field and always
has a non-zero net magnetic moment, eliminating possible silent
sources. However, the information obtained from the external
fields is still insufficient for imaging. For example, uniform
spherical magnetization distributions of different sizes can
produce the same magnetic dipole field. As a result, inversion
procedures using a 3-D grid of discrete dipoles
\cite{Kwiat,Sepulveda} are not unique.  It can be shown that at
least one magnetic field gradient has to be applied to solve the
inverse problem uniquely, as illustrated in Fig. 2. The magnetic
field gradient separates different slices of the sample in
frequency and within each slice an image can be uniquely obtained
from the array of sensors outside. This problem is analogous to
the determination of a 2-dimensional current density or
susceptibility distribution using magnetic field measurements
\cite{Roth,Tan,Sheltraw}. The inverse problem can be solved
exactly, however the spatial resolution drops exponentially with
the distance between the magnetometer plane and the imaging slice.
Thus, this imaging method would sacrifice some spatial resolution
to obtain very fast imaging speed.  With a large array of
sensitive detectors it should be possible to obtain a complete 3-D
image from a single FID in less than 1 msec, enabling new MRI
applications of time-varying processes.

\begin{figure}
\centerline{\includegraphics*[scale=0.5]{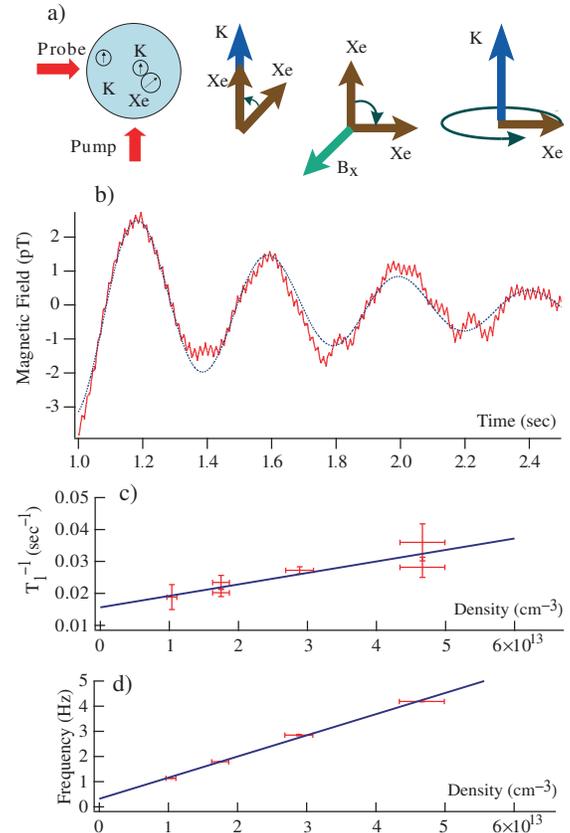}}
\caption{Detection of $^{129}$Xe NMR signal. a) Schematic of the
experiment and 3 main steps in the process: polarization of
$^{129}$Xe by spin exchange, tipping of $^{129}$Xe spins with a
constant field, and precession of $^{129}$Xe spins around the K
magnetization. b) A single $^{129}$Xe FID signal following the
tipping pulse at $t=0$ with a fit (dashed line). From the fit, the
initial amplitude of the signal at $t=0$ is 11.4 pT, the frequency
is 2.46 Hz, and the transverse relaxation time $T_2^{*}=
0.78$~sec. c) Measurement of $T_1$ as a function of K density. The
slope of the fit gives the K-Xe spin-exchange rate $\sigma_{\rm
SE} \bar{v}= 3.6(9)\times 10^{-16}$ s$^{-1}$cm$^{-3}$ and the
intercept gives the wall relaxation rate $T^{-1}_{1 \rm
wall}=0.016(3)\,{\rm s}^{-1}$. d) The $^{129}$Xe spin precession
frequency as a function of K density. The slope of the fit
8.4(3)$\times10^{-14}$ Hz/cm$^3$ is proportional to $\kappa_0$. }
\label{kxe}
\end{figure}

Another unique aspect of alkali-metal magnetometers is their
ability to interact directly with the nuclei of interest. This
interaction is particularly well understood in the case of noble
gases and has been used to produce hyperpolarized gases for a wide
range of applications \cite{WalkerRMP}. The attraction of the
alkali metal valence electron to the noble gas nucleus results in
an enhancement of the dipolar field created by the nuclear spins.
For a spherical cell, the effective field experienced by the K
atoms is given by
\begin{equation}
B_{\rm K}= \frac{8 \pi}{3} \kappa_0 M,
\end{equation}
where $M$ is the nuclear magnetization \cite{Schaefer}. Hence, the
classical dipolar field produced by nuclear spins is increased by
a factor of  $\kappa_0$ which ranges from  6 for $^3$He \cite{k0}
to about 600 for $^{129}$Xe \cite{Walker}.  The magnetization of K
atoms $M_{\rm K}$ also creates an effective field experienced by
the noble gas,
\begin{equation}
B_{\rm Xe}= \frac{8 \pi}{3} \kappa_0 M_K=\frac{8 \pi}{3} \kappa_0
g_s \mu_{\rm B} P_{\rm K} [{\rm K}]\label{BXe}
\end{equation}
In a high-density alkali-metal vapor this field causes $^{129}$Xe
atoms to precesses at a frequency of a few Hz while K atoms remain
in a nearly zero field.

In Fig.~\ref{kxe} a) we illustrate the principle of the
experiment. A small concentration of $^{129}$Xe atoms
(740~$\mu$Torr of $^{129}$Xe enriched to 80\%) is added to the
magnetometer cell. $^{129}$Xe is polarized parallel to the pump
beam by spin-exchange collisions with K atoms.
 To tip the $^{129}$Xe spins,  a static
transverse magnetic field $B_{x}$ of 1 mG is turned on for about
200 msec. The field causes K atoms to depolarize and $^{129}$Xe
atoms to precess by $\pi/2$. After the field is turned off, K
atoms are quickly repolarized and $^{129}$Xe precess around the
field $B_{\rm Xe}$. The transverse oscillating magnetization
generates the field $B_{\rm K}$ which is detected by the K
magnetometer.

Fig.~\ref{kxe}b) shows the FID signal of $^{129}$Xe nuclear
precession.  The data are well described by an
exponentially-decaying oscillation after subtracting a
slowly-varying background. The transverse spin relaxation time
$T_{2}^{*}$ is  determined by the inhomogeneities of the K
polarization across cell. We  found that applying a $B_{z}$ field
of 10-100~$\mu$G and increasing the optical pumping rate increases
the $^{129}$Xe signal due to a more uniform K polarization.

By measuring the equilibrium $^{129}$Xe signal from a train of
$\pi /2$ pulses as a function of the separation time between the
pulses we determined $T_{1}$ relaxation of $^{129}$Xe for
different temperatures corresponding to different densities of K.
Measurements of the effects of K-K spin-exchange collisions
\cite{Allred,Savukov} on the K Larmor resonance frequency and
linewidth were used to determined the density and the polarization
of K atoms. For example at 180$^{\circ}$C the potassium
polarization is $P_{\rm K}=85\%$ and the density is $[{\rm
K}]=2.9\times10^{13}$ cm$^{-3}$, about 3 times smaller than
saturated vapor pressure. The dependence of $1/T_{1}$ on the
density of K is shown in Fig.~\ref{kxe}c) from which we determine
the K-$^{129}$Xe spin-exchange cross-section, $\sigma_{\rm SE}=6.3
\times 10^{-21}$~cm$^{2}$, which compares well with a theoretical
estimate $\sigma_{\rm SE}=8 \times 10^{-21}$~cm$^{2}$
\cite{Walker}. We also measured the Xe precession frequency as a
function of K density, as shown in Fig.~\ref{kxe}d). In accordance
with Eq.~(\ref{BXe}) the frequency is proportional to the density
of K atoms. From  the slope of the fit we determine $\kappa_0
=540$, in good agreement with the theoretical estimate $\kappa_0
=660$ \cite{Walker}.

The equilibrium $^{129}$Xe polarization is given by $P_{\rm
Xe}=P_{\rm K}\sigma_{\rm SE} \bar{v}[{\rm K}]/(T^{-1}_{1 \rm
wall}+\sigma_{\rm SE} \bar{v}[{\rm K}])$ and is equal to
approximately 35\% at 180$^{\circ}$C. For the $^{129}$Xe density
of $2\times 10^{13}$~cm$^{-3}$ the effective field seen by K atoms
$B_K = 12$~pT, in excellent agreement with the measured signal of
11.4 pT after correcting for the signal decay during the dead time
of the magnetometer. The S/N  is approximately equal to $ 10$ in a
bandwidth of 10 Hz and the effective measurement volume determined
by the intersection of the pump and probe beams was about 1
cm$^3$. Thus, the magnetometer sensitivity is about $7\times
10^{11}/{\rm Hz}^{1/2}$ $^{129}$Xe atoms. With additional
optimization and better magnetic field shielding it should be
possible to achieve magnetic field sensitivity of better than 1
fT/Hz$^{1/2}$, giving sensitivity to about $10^9$ $^{129}$Xe spins
in a single shot. This detection technique  can be easily adapted
for detection of low $^{129}$Xe concentration in a flow-though
system \cite{Bastian} as long as $^{129}$Xe spends much less than
$T_1\sim 20 $ sec in the cell. $^{129}$Xe can be initially
polarized by optical pumping, flow through the sample where the
information is encoded in the longitudinal polarization and then
flow through the K cell for detection. This work was supported by
NSF, Packard Foundation and Princeton University.

\end{document}